# Method Chunks Selection by Multicriteria Techniques: an Extension of the Assembly-based Approach

Elena Kornyshova[1,2], Rébecca Deneckère[1], and Camille Salinesi[1]

1  CRI, University Paris 1 – Panthéon Sorbonne
90, rue de Tolbiac, 75013 Paris, France,

2  ECD, Saint-Petersburg State University of Economics and Finance
21, Sadovaia Str, 191023 Saint-Petersburg, Russia
{elena.kornyshova,rebecca.deneckere,camille}@univ-paris1.fr,
WWW home page: http://crinfo.univ-paris1.fr/

**Abstract**. The work presented in this paper is related to the area of situational method engineering (SME). In this domain, approaches are developed accordingly to specific project specifications. We propose to adapt an existing method construction process, namely the assembly-based one. One of the particular features of assembly-based SME approach is the selection of method chunks. Our proposal is to offer a better guidance in the retrieval of chunks by the introduction of multicriteria techniques. To use them efficiently, we defined a typology of projects characteristics, in order to identify all their critical aspects, which will offer a priorisation to help the method engineer in the choice between similar chunks.

## 1  Introduction

It is now clearly assumed that one development process cannot fit all the existing problems and development contexts. This assumption has lead to the development of the Method Engineering domain, and more particularly of Situational Method Engineering (SME) [1] [2]. In this domain, approaches have been developed to adapt existing methods to deal with the specifications of the project at hand. It allows the construction of a specific process to meet the requirements of each particular situation by reusing and assembling parts of existing methodologies called either fragments [3], chunks [4], patterns [5], etc, that, similarly to a software component, can be treated as separated unit. The knowledge encapsulated in these small method parts is generally stored in a classic library repository called Method Base [4] [6] [7].

Following a complete assembly SME approach consists of executing the following phases: (a) identification and formalisation of the method chunks,

(b) storage in a method chunks base, (c) chunks selection following the project needs, and (d) assembling of the selected method chunks. In this paper, we will consider the SME aspect regarding the selection of chunks in the repository. In our proposal, we refer to the notion of "chunk" to describe every type of small method parts (considered also as fragment or as pattern). The problem of chunk retrieval is an important part of this process and has to be easy and effective.

The assembly based approach [8] uses a process (assembly process model – APM) that guides the engineer in the elaboration of a requirement map and uses this map in order to select a set of related chunks. The final selection is then realised with the help of similarity measures inspired from those proposed by [9] and [10]. They distinguish two types of measures: those which allow to measure the similarity of the elements of *product models* and those which allow to measure the closeness of *process models* elements.

The similarity measures are provided in order to compare the method requirements with the solutions proposed by the selected chunks but their application is difficult. First, the difference between the formulation of requirements to achieve and of requirements that can be achieved is more or less inexistent, which made the requirements map creation difficult. Second, the results obtained by an application of the similarity measures are not simple to handle. Furthermore, the cost of a project can increase as, in order to offer a good comparison, method engineers have to manage an increasing number of artefacts, which induce a combinatory explosion of all the values to calculate. Finally, even if all these issues are solved, the final selected chunks may be quite similar; this means that the method engineer has to choose one over the other and to discriminate between them.

To solve these difficulties, we propose an extension of the APM by the introduction of multicriteria (MC) techniques (or MC methods). Our objectives are to (a) guide chunk retrieval and (b) to propose a priorisation of the selected chunks in order to guide the method engineer into the final selection process. In order to use the full potentiality of the MC techniques, we also propose a project characteristics typology, in order to identify all its critical aspects. This typology is an adaptation of two similar works. The first one is the typology created by Kees Van Slooten and Bert Hodes in [11] to prove that the project approach is affected by the project context. The second was made by Isabelle Mirbel and Jolita Ralyté in [12]. In this work, they define the concept of Reuse frame and they apply it to the assembly approach. Their reasons are threefold: (a) to help the chunk selection by better qualifying them, (b) to enable the use of more powerful matching techniques to retrieve them when looking at similar methodological problems and (c) to express better methodological needs for a specific project, improving this way the chance to get adequate and useful method chunks. The merging of these two existing typologies and their adaptation to be used by MC techniques will multiply the process efficiency.

Our approach is presented in this paper as follows: In the section 2, we give a brief introduction in MC techniques. The section 3 describes the assembly-based approach extended by MC techniques with an example. The section 4 presents conclusion and future works.

## 2   Multicriteria Techniques

Multicriteria techniques currently dominate in the field of decision-making [13], [14]. They appeared at the beginning of the Sixties, and their number and application contexts increase continually. For example, these techniques are employed for requirements priorisation [15], to choose evolution scenario [16], or to make operational decisions [17].

Generally, a decision-making problem is defined by the presence of alternatives. The traditional approach consists in using only one criterion to carry out the selection between alternatives. The traditional example is the selection of the projects according to the net present value (NPV). However, using a single criterion is not sufficient when the consequences of the alternatives to be analyzed are important [18].

The goal of the multicriteria decision-making (MCDM) techniques consists in defining priorities between alternatives (actions, scenarios, projects) according to multiple criteria. In contrast to monocriterion approach, MC techniques allow a more in-depth analysis of problem because of taking into consideration various aspects. Nevertheless, their application has proved more difficult.

In spite of their complexity, MC techniques are often chosen and used by companies. In general, the MC formulation of a problem is based on the definition of [19]:
 – alternatives set represented by "concurrent" actions,
 – criteria (attributes) set defined by parameters to be considered for priorisation,
 – alternatives evaluations according to criteria (partial evaluations, which are obtained by assignment of values to each alternative according to all criteria),
 – aggregation rules (to select an alternative, it is necessary to incorporate the partial evaluations in a general evaluation). The aggregation rules differ in different techniques.

According to this, the decision-making steps are defined as follows:
1. diagnostics of problem (necessity to define priorities),
2. identification of problem's parameters: alternatives, criteria,
3. alternatives partial estimations,
4. priorities definition.

Five families of MCDM techniques can be considered: MAUT [20], AHP [21], outranking techniques [18], weighting techniques [22], and fuzzy techniques [23]. These are not detailed here for the sake of space.

## 3   Extended Assembly-based Approach

Using MC techniques allow to integrate new parameters into method chunk selection. We propose to adapt namely the assembly based SME approach by integrating of MC techniques expression.

The basic and extended APM are illustrated in Fig. 1 using the MAP formalism [24].

The intentional modelling of MAP provides a generic model based on intentions (goals) and the possible strategies to achieve each intention. The map is presented as a graph where nodes are *intentions* and edges are *strategies*. The directed nature of the graph shows which intentions can follow which one. An edge enters a node if its manner can be used to achieve its intention. Since there can be multiple edges entering a node, the map is able to represent the many manners that can be used for achieving an intention. The map includes two predefined intentions: "Start" and "Stop", which mean accordingly the beginning and the end of the process. An important notion in process maps are the *sections* witch represent the knowledge encapsulated in a triplet <source intention, strategy, target intention>, in other terms, the knowledge corresponding to a particular process step to achieve an intention (the target intention) from a specific situation (the source intention) following a particular technique (the strategy).

In the following figure, the basic components of APM are presented by solid lines, and the components proposed to extend the basic approach are exposed by dashed lines.

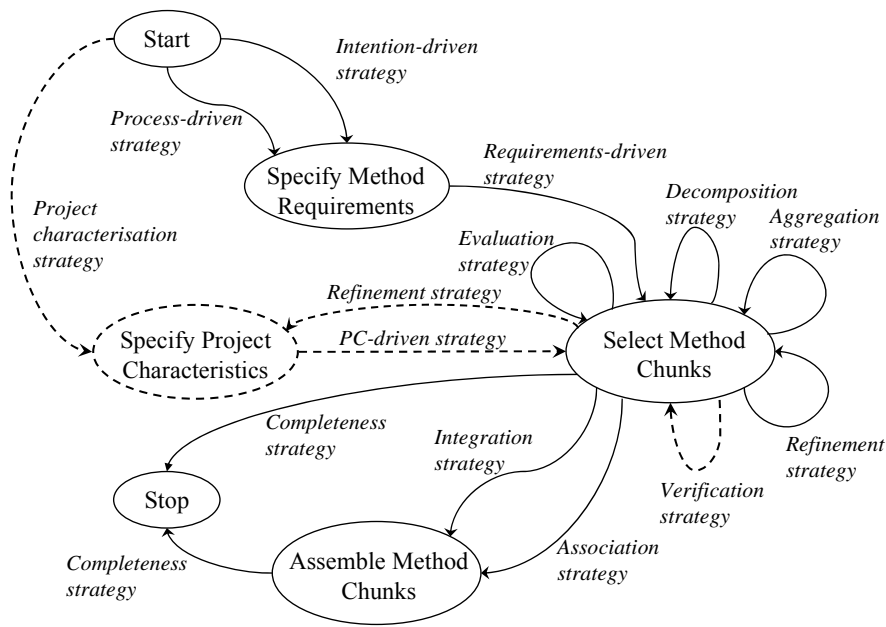

**Fig. 1.** Basic and extended APM

This map is described in the following sections. Firstly, we present the basic APM, secondly, the extended one, and, finally, an illustrative example.

### 3.1  Basic Assembly-based Approach

The APM [8] is based on the notion of "chunk" as a representation of a method small unit. It proposes different ways to select them that match requirements as well as different strategies to assemble them. It is based on the achievement of two key intentions: *Select method chunks* and *Assemble method chunks*. Achieving the first intention leads to the selection of chunks from the method base that matches the requirements. The second intention is satisfied when the selected chunks have been assembled in a consistent manner.

The process starts by selecting candidate chunks that are expected to match the requirements expressed in a requirements map. Guidelines suggest formulating queries to the method base in order to identify the chunks that are expected to match part or the totality of the requirements. A set of strategies (*decomposition, aggregation, refinement, decomposition, aggregation*) help to refine the candidate chunk selection, but, any time a chunk has been retrieved, it can be validated by applying an *evaluation strategy*. This helps in evaluating the degree of matching of the candidate chunk to the requirements. This is based on similarity measures between the requirements map and the map of the selected chunk.

When at least two chunks have been selected, the method engineer can progress to the assembly of these chunks. Two strategies, namely the *integration strategy* and the *association strategy*, are proposed to fulfil the intention *Assemble method chunks*. The choice of the strategy depends on the presence/absence of overlaps between the chunks to assemble. Similarity measures are used to compare chunks before their assembly and to identify whether they are overlapping. This will help to choose the right strategy between the *integration strategy* and the *association strategy*.

### 3.2 Proposed Extension of Assembly-based Approach

As we can see in Fig. 1, the basic APM may be extended by the following sections:
1. Specify Project Characteristics by Project characterisation strategy,
2. Specify Project Characteristics by Refinement strategy,
3. Select Method Chunks by Project Characteristics (PC)-driven strategy,
4. Select Method Chunks by Verification strategy.

These sections are described in the following paragraphs according to two intentions: *"Specify Project Characteristics"* and *"Select Method Chunks"*.

### 3.2.1 Specify Project Characteristics

Project characteristics influence method chunks selection. Each method chunk is described according to its contribution to these characteristics. This typology can be enriched by introduction of characteristics proper to concrete methods (such a used approach, tool presence, notation, difficulty etc).

*Project characteristics typology*
Project characteristics describe the main properties of IS development project. Their difference with method requirements of basic APM lies in the way of definition and presentation. The method requirements are analysed and expressed in the form of requirements map, whereas the project characteristics form a predefined typology that method engineer investigates in order to choose those, which are needed for a project.

Based on studies [11] [12], we propose a typology of project characteristics, which includes four dimensions: organisational, human, application domain, and development strategy.

The typology of project characteristics is illustrated on Tables 1, 2, 3, and 4. The characteristics proposed in this table are either inspired from the works presented in [11] and [12] or suggested in this paper. In order to differentiate them in the table, we identify the source (1) as the work of Van Slooten [11], the source (2) as Mirbel's [12] and ours will be noted as the source (3).

The organisational dimension highlights organisational aspects of IS development project and includes the following characteristics: management commitment, importance, impact, time pressure, shortage of resources, size, and level of innovation (Table 1).

**Table 1.** Organisational dimension.

| Characteristic | Values | Source |
|---|---|---|
| Management commitment | {low, normal, high} | (1), (2), (3) |
| Importance | {low, normal, high} | (1), (3) |
| Impact | {low, normal, high} | (1), (2), (3) |
| Time pressure | {low, normal, high} | (1), (2), (3) |
| Shortage of resources | {low, normal, high} | (1), (2), (3) |
|  | {human, means} | (1), (2) |
|  | {financial resources, human resources, temporal resources, informational resources} | (3) |
| Size | {low, normal, high} | (1), (2), (3) |
| Level of innovation | {low, normal, high} | (1), (2), (3) |
|  | {business innovation, technology innovation} | (2), (3) |

The human dimension describes the qualities of persons involved into IS development project. It includes the following characteristics: resistance and conflict, expertise, requirements clarity and stability, user involvement, stakeholder number (Table 2).

**Table 2.** Human dimension

| Characteristic | Values | Source |
|---|---|---|
| Resistance and conflict | {low, normal, high} | (1), (3) |
| Expertise (knowledge, experience, and skills) | {low, normal, high} | (1), (2), (3) |
|  | {tester, developer, designer, analyst} | (2), (3) |
| Clarity and stability | {low, normal, high} | (1), (2), (3) |
| User involvement | {real, virtual} | (2), (3) |
| Stakeholder number | num | (3) |

The application domain dimension includes formality, relationships, dependency, complexity, application type, application technology, dividing project, repetitiveness, variability, and variable artefacts (Table 3).

**Table 3.** Application domain dimension.

| Characteristic | Values | Source |
|---|---|---|
| Formality | {low, normal, high} | (1), (2), (3) |
| Relationships | {low, normal, high} | (1), (3) |
| Dependency | {low, normal, high} | (1), (2), (3) |
| Complexity | {low, normal, high} | (1), (3) |
| Application type | {intra-organization application, inter-organization application, organization-customer application} | (2), (3) |
| Application technology | {application to develop includes a database, application to develop is distributed, application to develop includes a GUI} | (2), (3) |
| Dividing project | {one single system, establishing system-oriented subprojects, establishing process-oriented subprojects, establishing hybrid subprojects} | (1), (2), (3) |
| Repetitiveness | {low, normal, high} | (3) |
| Variability | {low, normal, high} | (3) |
| Variable artefacts | {organisational, human, application domain, and development strategy} | (3) |

The development strategy dimension gathers source system, project organization, development strategy, realization strategy, delivery strategy, tracing project, and goal number (Table 4).

**Table 4.** Development strategy dimension.

| Characteristic | Values | Source |
|---|---|---|
| Source system | {code reuse, functional domain reuse, interface reuse} | (2), (3) |
| | {weak, medium, strong} | (2), (3) |
| Project organization | {standard, adapted} | (1), (2), (3) |
| Development strategy | {outsourcing, iterative, prototyping, phase-wise, tile-wise} | (1), (2), (3) |
| Realization strategy | {at once, incremental, concurrent, overlapping} | (1), (2), (3) |
| Delivery strategy | {at once, incremental, evolutionary} | (1), (2), (3) |
| Tracing project | {weak, strong} | (1), (2), (3) |
| Goal number | {one goal, multi-goals} | (3) |

*Specify Project Characteristics by Project characterisation strategy*
This section consists in the identification of characteristics for a given project. The method engineer explores the project characteristics typology and brings out the project critical aspects, which are crucial for the current project.

*Specify Project Characteristics by Refinement strategy*
The refinement strategy is similar to this one of the basic APM. The distinction is concluded in a refinement objective. The selection result may be presented by a set of method chunks, which are homogeneous, i.e. have the same description with regard to previously identified project characteristics. Then, additional information is required to define more precisely the differences between homogeneous method chunks. In this case, the refinement aims to specify more closely the project characteristics.

### 3.2.2 Select Method Chunks

*Select Method Chunks by PC-driven strategy*

The PC-driven strategy consists in application of MC techniques for selecting alternatives method chunks.

This section can be itself refined by a process map (illustrated on Fig. 2), which contains two main intentions: *"Define weights"* and *"Define priorities"*.

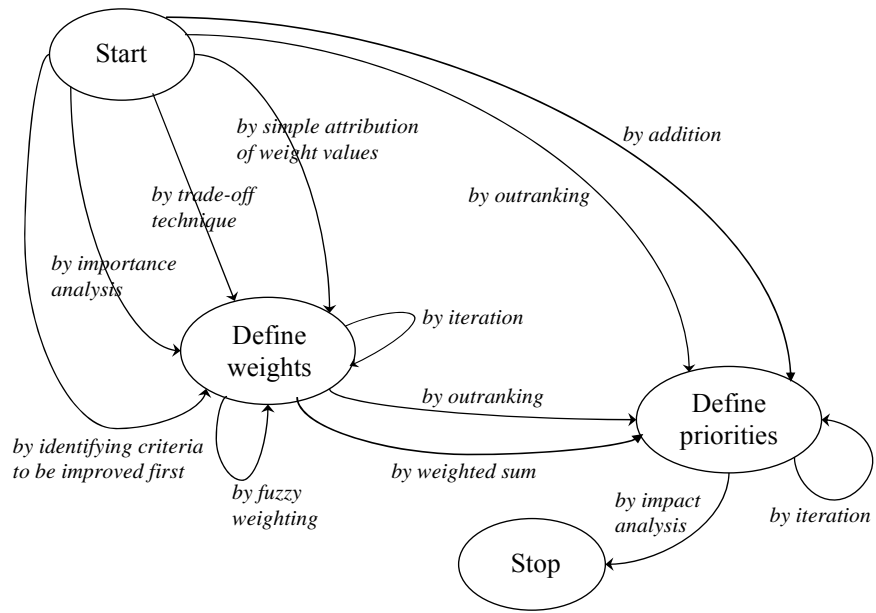

**Fig. 2.** Select Method Chunks by PC-driven strategy's process map

The choice between these two intentions is made according to needs for criteria weighting. Criteria weighting enables to analyse their relative importance. When they are not weighted, it means that their relative importance is equal.

These two strategies are developed in the following paragraphs (for more details on outranking and weighting techniques, see Appendix 1).

1. *Define weights.* This intention can be achieved *"by simple attribution of weight values"*, *"by identifying criteria tot be improved first"* (SWING), *"by trade-off technique"* (trade-off weighting), and *"by importance analysis"* (SMART). The choice between these possibilities can be carried out in function of decision-maker preferences. This intention can be achieved *"by iteration"* (when the result must be specified) and *"by fuzzy weighting"* (when fuzzy values are needed).

2. *Define priorities* (Priorisation). There are two ways to achieve this intention that are: priorisation strategy with or without weighting and priorisation strategy with weighting. The objective of this stage is to aggregate the alternatives evaluations into a unique (aggregated) evaluation and to priory alternatives.

*Priorisation strategy without weighting.* To carry out this strategy, we suggest application of the following strategies: *"by outranking"* (outranking) without weighting or *"by addition"*. The addition of values requires that all of them must have a homogeneous qualitative nature and be normalised. The outranking can be applied to all data types (quantitative and qualitative) and does not require the normalisation. However, it is most complicated.

*Priorisation strategy with weighting.* The possible strategies are *"by outranking"* (outranking) with weighting or *"by weighted sum"* (weighting techniques). .The difference between the given strategies is similar to the previous selection.

This intention too can be completed *"by iteration"* if the result has to be specified.

The section *Stop by impact analysis* allows analysing the results of priorisation by considering the impact and interactions between selected chunks.

Hence, we have identified four main strategies corresponding to these four different MC techniques. Arguments for choosing one of them are presented in Table 5.

**Table 5.** Arguments for choosing a main strategy.

|  | **Addition** | **Outranking** |
| --- | --- | --- |
| **Without weighting** | All criteria have the same relative importance; All criteria have a homogeneous qualitative nature and are normalised. | All criteria have the same relative importance; All data types. |
| **With weighting** | The criteria have different relative importance; All criteria have a homogeneous qualitative nature and can be normalised. | The criteria have different relative importance; All data types. |

This table presents the combination of arguments allowing the user to choose the right MC technique. The arguments include two essential aspects being the relative importance of the criteria ("same" or "different") and their nature ("all" or "homogeneous qualitative and normalised").

*Select Method Chunks by Verification strategy*
This strategy aims at verifying adequacy of chunks selected by MC techniques: if the result is not sufficient, other project characteristics are needed for final decision-making. Then the section *"specify project characteristics by refinement"* is available.

### 3.3 Example

To illustrate our proposal, we have selected method chunks that deal with information system (IS) security within requirements engineering (RE).

Five chunks of RE methods designed for analysing IS security were identified: NFR Framework [25], KAOS [26], Secure Tropos [27], GBRAM [28], and Misuse Cases [29]. The comparison of these methods is presented in [30]. Within this example, we illustrate only one part of extended APM that concerns the application of MC techniques in SME.

The given project is described by:
- the great influence on the whole organisation;
- the need for ensuring the greater progress;
- the organisation does not have the experts in this field and does not plan to employ them;
- the need for a better explanation of method chunks and their application.

The method engineer has chosen three project characteristics (since the weights of the others are equal to zero) and has described the method chunks according to methods properties. Thus, these methods chunks are compared according to six criteria, which concern two groups: project characteristics and proper method characteristics. The first group includes impact, level of innovation, and expertise. The second group comprises guidance, approach, and formalism.

Depending on project description, the method engineer has defined the following preferences rules for these criteria:
- *Impact on organisation:* maximum;
- *Level of innovation:* maximum;
- *Required expertise:* minimum;
- *Guidance:* a predefined taxonomy is better than heuristics, which is better than a simple guidelines;
- *Approach:* a systemic approach is better than exploratory, which is better than explanatory.
- *Formalism:* a formal approach is better than semi-formal one, which is better than informal one.

The summary of chunks evaluation is presented in Table 6.

**Table 6.** IS security chunks evaluation.

| Criteria | NFR Framework | KAOS | Secure Tropos | GBRAM | Misuse Cases |
|---|---|---|---|---|---|
| *Project Characteristics* | | | | | |
| Impact | high | low | high | low | normal |
| Level of innovation | high | high | low | high | high |
| Expertise | normal | high | high | normal | low |
| *Method Chunk Characteristics* | | | | | |
| Guidance | predefined taxonomy | reuse of generic refinement patterns, heuristics | *No guidance* | documents analysis, heuristics | guidelines |
| Approach | explanatory | exploratory | systemic | *Not applicable* | explanatory |
| Formalism | semi-formal | formal | formal | informal | informal |

In order to compare these chunks and to select one of them, which is more adapted to the given project, we have applied three different calculations: simple addition, weighted sum, and outranking with weighting.

1) The *simple addition* was applied to the first three criteria, which are *"quantifiable"*. Two method chunks, which are the best ones, present the result: NFR Framework and Misuse Cases (See Table 7).

**Table 7.** Method selection with simple addition.

| Criteria | NFR Framework | KAOS | Secure Tropos | GBRAM | Misuse Cases |
|---|---|---|---|---|---|
| Impact | 3 | 1 | 3 | 1 | 2 |
| Level of innovation | 3 | 3 | 1 | 3 | 3 |
| Expertise | 2 | 1 | 1 | 2 | 3 |
|  | **8,00** | 5,00 | 5,00 | 6,00 | **8,00** |

2) In the case of *weighted sum*, we add the weights assigned to criteria. These weights are defined by importance analysis (Appendix 1). The chunk "Misuse Cases" is the best one (see Table 8).

**Table 8.** Method selection with weighted sum.

| Criteria | Weights | NFR Framework | KAOS | Secure Tropos | GBRAM | Misuse Cases |
|---|---|---|---|---|---|---|
| Impact | 0,30 | 3 | 1 | 3 | 1 | 2 |
| Level of innovation | 0,20 | 3 | 3 | 1 | 3 | 3 |
| Expertise | 0,50 | 2 | 1 | 1 | 2 | 3 |
|  |  | 2,50 | 1,40 | 1,60 | 1,90 | **2,70** |

3) To apply the outranking technique, we selected ELECTRE [18] (Appendix 1). All calculations are not presented here for the sake of space. The concordance and discordance matrices developed in our case study are shown in Table 9 (Table 9.a – concordance matrix; Table 9.b – discordance matrix). Application of outranking techniques allows considering the last three criteria, which are not quantifiable.

**Table 9.** Method selection with outranking.

a)

|  | Fr1 | Fr2 | Fr3 | Fr4 | Fr5 |
|---|---|---|---|---|---|
| **Fr1** | X | 0,45 | 0,45 | 0,85 | 0,85 |
| **Fr2** | 0,60 | X | 0,50 | 0,85 | 0,75 |
| **Fr3** | 0,65 | 0,80 | X | 0,65 | 0,65 |
| **Fr4** | 0,35 | 0,60 | 0,50 | X | 0,35 |
| **Fr5** | 0,60 | 0,30 | 0,35 | 0,70 | X |

b)

|  | Fr1 | Fr2 | Fr3 | Fr4 | Fr5 |
|---|---|---|---|---|---|
| **Fr1** | X | 0,33 | 0,67 | 0,50 | 0,50 |
| **Fr2** | 1,00 | X | 1,00 | 0,50 | 1,00 |
| **Fr3** | 1,00 | 0,67 | X | 0,67 | 1,00 |
| **Fr4** | 1,00 | 0,67 | 1,00 | X | 0,33 |
| **Fr5** | 0,67 | 1,00 | 0,67 | 1,00 | X |

As we can see, only one alternative (NFR Framework) dominates the others without any particular shortcoming in terms of discordance. As a result, this first chunk is selected.

Application of different MC techniques for selecting method chunks gives different results. The simple addition is the simplest technique, but it implies the following disadvantages: a) it does not take into account the relative importance of criteria and b) it is applicable only for numeric or easy quantifiable criteria. The weighted sum supports the criteria relative importance, but saves the restrictions on data type (quantitative). The outranking technique is more complex, its application requires additional skills. Nevertheless, the result is defined more precisely with consideration of all data types. In this case, the chunk "NFR Framework" was selected in order to analyse the requirements of IS security.

Hence, the approaches using diverse MC techniques imply a selection of different method chunks. For this reason, we recommend to use one of strategies described above to specify and to select the method chunks (by addition, by weighting, or by outranking) according to available criteria.

## 4   Conclusion

We have proposed an adaptation of the existing assembly process with the introduction of MC techniques. The two approaches (basic and extended) may be combined within the same method engineering process as it will offer a more complete guidance to select chunks.

Our objective is twofold. Firstly, we offer the possibility to the method engineer to qualify the method chunks by their correspondence with projects and to choose between similar chunks by an application of MC techniques. Secondly, we propose to characterise the project and the chunks to improve their selection. This typology allows to identify all their critical aspects and to weight them. Within our example, we showed the utility of application of MC techniques and revealed that different MC techniques give different selection result.

In near future, our research perspectives include:
- improve the guidance;
- adapt other situational methods by integrating MC techniques;
- improve the typology presented in this paper in order to take into account other critical characteristics;
- extend the MC techniques application to the field of System Engineering based on MC techniques chunks.

## Appendix 1

This appendix presents a brief description of two groups of MC techniques: outranking and weighting techniques.

*Outranking techniques*

Outranking techniques [17], [18], [31] are inspired from the theory of social choice [17]. There are two kinds of approaches in the family of outranking techniques: ELECTRE (created by Roy, since 1968) and PROMETHEE (created by Brans J.P., Mareschal B., and Vincke Ph, since 1984) [18], [31]. The most known technique is ELECTRE (ELimination Et Choix Traduisant la REalité, B. Roy / Elimination And Choice Corresponding to Reality). Outranking techniques serve for approaching complex choice problems with multiple criteria and multiple participants. Outranking indicates the degree of dominance of one alternative over another. Outranking techniques enable the utilization of incomplete value information and, for example, judgments on ordinal measurement scale.

It includes the following steps:

1. Calculation of the indices of concordance and discordance on the basis of estimation of two given alternatives. These indices define the concordance and discordance following the assumption that alternative A is preferred to alternative B. The principle is that the decision maker estimates that alternative A is at least as good as B if the majority of the attributes confirm it (concordance principle) and the other attributes (minority) are not strong enough (discordance principle).

2. Definition of levels for the concordance and discordance indices. If the concordance index is higher then defined level and the discordance one is lower, then an alternative is preferred to the other. If it is note the case, alternatives are incompatible (what means that A is preferred to B according to criterion X, and B is preferred to A according to the criterion Y).

3. Elimination of dominated alternatives. Then a first alternatives subset is obtained, which can be either equivalent, or incompatible.

4. Iterative application of stages 2 and 3 with "lower" levels of concordance and discordance indices. A more restricted subset of alternatives is then carried out.

The procedure is applied until a suitable subset is obtained. A last subset includes the best alternatives. The order of the obtained subsets determines the alternatives scale according to their criteria given suitability.

The ELECTRE family has several members: ELECTRE I (for choice problems), ELECTRE II, ELECTRE III, ELECTRE IV (for ranking problems), ELECTRE TRI (for alternatives sorting). An advantage of outranking techniques is that they are based on step-by-step identification of decision makers' preferences. A detailed analysis makes it possible to the decision makers to formulate his preferences and to define compromises between the criteria. The incompatibility relation can be employed to find the contradictory pairs of alternatives, to stop on a subset whose choice is justified (with available information). Difficulties can appear during the weight definition by the decision maker. Moreover, the appearance of the cycles (when alternative A is preferred to B, B is preferred to C and C is preferred to A) is rare but is not excluded.

*Weighting techniques*

Weighting techniques include SMART (Simple Multiattribute Technical Rating), SWING, and Trade-off weighting [22], [32], and [33]. They are characterised by a weight assignment to the decision criteria. Aggregation of the evaluations is based on weighted sum.

The SMART technique (proposed by W. Edwards), which appeared the first, includes the following stages: criteria scaling according to their importance, criteria attribution of a value from 1 to 100, calculation of the relative importance of each criterion. We call it definition of criteria weights by importance analysis.

In SWING weighting (D. Winterfeldt и W. Edwards), all criteria are supposed bad. The expert chooses the one, which must be improved firstly and a value of 100 is attributed to this criterion. The same operation is carried out with the other criteria to determine their values (by identifying criteria to be improved first).

In Trade-off weighting (H. Raiffa and R.L. Keeney) the decision maker compares two hypothetical alternatives according to two criteria; other criteria are invariable. The weights of these two criteria are refined so that the values of two given weighted alternatives have the same importance for the decision maker. This operation is repeated until all the weights are defined.